\documentclass{article}
\usepackage[T1]{fontenc} 
\usepackage[utf8]{inputenc} 
\usepackage{times}
\usepackage[scaled=0.85]{helvet}
\usepackage{graphicx}
\usepackage{ifthen}
\usepackage{xspace}
\usepackage{alltt}
\usepackage{latexsym}
\usepackage{url}
\usepackage{amssymb}
\usepackage{amsfonts}
\usepackage{amsmath}
\usepackage{stmaryrd}
\usepackage{enumerate}
\usepackage{cite}
\usepackage[colorlinks=true,pdfstartview=FitV,linkcolor=blue,citecolor=blue,urlcolor=blue]{hyperref}
\usepackage{xspace}

\newboolean{showcomments}
\setboolean{showcomments}{true}
\ifthenelse{\boolean{showcomments}}
  {\newcommand{\bnote}[2]{
        \fbox{\bfseries\sffamily\scriptsize#1}
    {\sf\small$\blacktriangleright$\textit{#2}$\blacktriangleleft$}
   }
   
  }
  {\newcommand{\bnote}[2]{}
   
  }

\newcommand{\commented}[1]{}

\newcommand{\eg}{\emph{e.g.,}\xspace}
\newcommand{\ie}{\emph{i.e.,}\xspace}

\newcommand{\ct}[1]{{\textsf{#1}}\xspace}

\newenvironment{code}
    {\begin{alltt}\sffamily}
    {\end{alltt}\normalsize}

\usepackage{url}
\makeatletter
\def\url@leostyle{%
  \@ifundefined{selectfont}{\def\UrlFont{\sf}}{\def\UrlFont{\small\sffamily}}}
\makeatother
\newcommand{\window}{\textsf{window}\xspace}
\newcommand{\this}{\textsf{this}\xspace}

\newcommand{\js}{JavaScript\xspace}
\newcommand{\safejs}{SafeJS\xspace}
\newcommand{\dom}{DOM\xspace}
\newcommand{\vdom}{\safejs VDOM\xspace}

\newcommand{\requirejs}{RequireJS\xspace}
\newcommand{\jsdom}{jsdom\xspace}
\newcommand{\contextify}{Contextify\xspace}
\newcommand{\nodejs}{node.js\xspace}

\begin{document}

\title{\safejs: Hermetic Sandboxing for \js}
\author{Damien Cassou \and St\'ephane Ducasse \and Nicolas Petton}
\date{}
\maketitle

\begin{abstract}
Isolating programs is an important mechanism to support more secure applications. Isolating program in dynamic languages such as \js is even more  challenging since reflective operations can circumvent simple mechanisms that could protect program parts. In this article we present \safejs, an approach and implementation that offers isolation based on separate sandboxes and control of information exchanged between them.

In \safejs, sandboxes based on web workers do not share any data. Data exchanged between sandboxes is solely based on strings. Using different policies, this infrastructure supports the isolation of the different scripts that usually populate web pages. A foreign component cannot modify the main \dom tree in unexpected manner.

Our \safejs implementation is currently being used in an industrial
setting in the context of the Resilience FUI 12 project.

\end{abstract}

\section{Introduction}

A mashup is a web page aggregating multiple sources of information. In
a \js-based mashup, such a source of information often requires
including an external script in the hosting web page for data
retrieval or UI building. Including a \js script from an external
source into a web page raises security concerns, as well as unresolved
questions about secure communication between the script and the
hosting web page.

\js semantics is permissive with respects to security
\cite{Maff09a,Guha10a,Rich11a}. Several attempts to isolate \js
programs have been made : Caja \cite{Mill08a}, FBJS\footnote{a subset
  of JavaScript for Facebook}, AdSafe \cite{Poli11a}. These solutions
either use a subset of \js or rely on code rewriting.

Our approach is based on separate thread-based local sandboxes and
controlled policies to access the main environment. On this aspect,
the idea is close to the one of Software Isolation Processes (SIP) of
Singularity OS \cite{Hunt07a}. \safejs is available for download under a free open-source
license.\footnote{\url{https://gitorious.org/safe-js/safe-js}}

The contributions of this technical report are:

\begin{itemize}
\item strong isolation of \js based on web workers ;
\item fully working implementation used in an industrial setting.
\end{itemize}

\section{DOM and Web Workers}

This section presents some web standards (\dom and web workers) that
are key to \safejs design and implementation. The \dom is the document
tree of any XML and HTML document. The DOM is virtually used by all
web applications. The web worker is an infrastructure to run isolated
processes in a web page. The web worker draft specification is already
implemented by modern web browsers.

\paragraph{\dom.} The W3C defines the \emph{\dom} (Document Object
Model) as ``\emph{an application programming interface (API) for HTML
  and XML documents}''\cite{DOM98}. The \dom specification consists of
3 stacked levels, each level building on the level below. The \dom
Level 1, provides an API to let developers add, modify and delete
elements from an HTML (in fact any XML) document. The \dom represents
any HTML document like a tree of node objects, each object
representing a particular tag of the document. In its Level 2, the
\dom additionally specifies how developers can manipulate other parts
of HTML documents such as CSS and mouse events. The \dom Level 3 adds
other APIs such as a keyboard event handling API and an XML
serialization API. \js applications manipulate the DOM (\eg adding and
removing nodes) to impact the visual representation of the web page.

\paragraph{Web Workers.} The W3C is in the process of specifying a
\emph{worker} API that ``\emph{allows Web application authors to spawn
  background workers running scripts in parallel to their main
  page}''\footnote{\url{http://www.whatwg.org/specs/web-apps/current-work/multipage/workers.html}}.
This draft specification enforces that the code executing within a
worker is completely isolated from the main thread and other workers:
a worker can only communicate through string-based messages, \ie a
worker has no access to the \dom API and thus can not modify the web
document. In the following, we will talk about the ``worker
specification'' even though this document is still in a draft stage
and subject to change. Still, this document is sufficiently advanced
to have compatible implementations in all modern web browsers.

The idea behind \safejs is to run each unsafe script inside its own
worker and give this worker a virtual \dom for the script to play with.
This virtual \dom is compatible with the standard \dom but restricts
which changes impact the document.

\section{\safejs runtime overview}

\begin{figure}
  \centering
  \includegraphics[width=\textwidth]{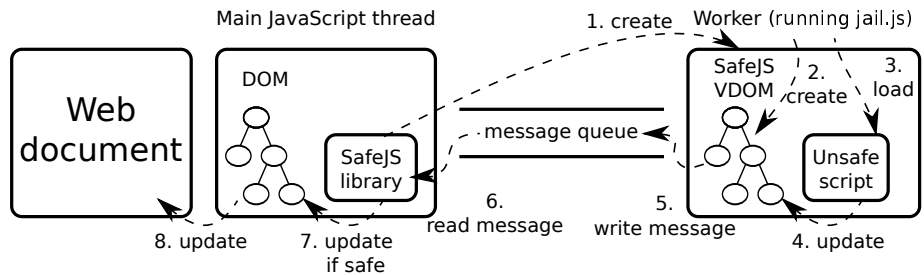}
  \caption{High-level architecture of \safejs}
  \label{fig:global-architecture}
\end{figure}

Figure \ref{fig:global-architecture} summarizes the high-level
architecture of \safejs. A standard web browser provides a view of the
web document, a main \js thread, and a \dom (accessible from the main
\js thread). The web developer must include the \safejs
library inside the document and pass each unsafe script he wants to
embed in the document to \safejs. As soon as an unsafe script is
passed to \safejs, \safejs creates a dedicated \js worker (step
\ct{1.}). This worker first creates a \safejs Virtual \dom (known as the \vdom)
mimicking the main \js thread's \dom (step \ct{2.}). The worker then
loads and starts the unsafe script (step \ct{3.}). The unsafe script
can freely read from the \vdom. The unsafe script can also update the
\vdom (step \ct{4.}). When this happens, the \vdom writes a message to
the message queue (step \ct{5.}). The \safejs library then reads the
message at the other end of the queue (step \ct{6.}). When \safejs
detects a forbidden update, \safejs will either ignore the request or
kill the worker and its malicious script; this choice depends on the
\safejs configuration. If the update is authorized, \safejs performs
the equivalent update on the real \dom (step \ct{7.}). The web browser
finally updates the web page (step \ct{8.}).

We now detail each step.

\subsubsection*{Step 0. -- Initializing \safejs}

The web developer must first load the \safejs library through
\requirejs:\footnote{\url{http://requirejs.org/}}

\begin{code}
<script type='text/javascript' data-main="safe" src='js/require.js' />
\end{code}

The \ct{data-main} declaration instructs \requirejs to load \safejs.
The next step for the web developer is to pass each unsafe script to
\safejs:

\begin{code}
<script type='text/safe-javascript' src='js/spy.js' node='#unsafeDiv' policy='read-write' />
\end{code}

This declaration instructs \safejs to load the unsafe \ct{js/spy.js}
script. This script can be any \js code compatible with client-side
usage. The \ct{type} parameter value must be set to
\ct{text/safe-javascript}: this prevents the web browser from loading
the script in the main thread and this indicates to \safejs that the
script must be sandboxed. The \ct{node} parameter indicates the HTML
target node which will be made accessible from the unsafe script: in
this case, the HTML node with the \ct{unsafeDiv} id. This target node
will become the only node visible from the \ct{spy.js} unsafe script.
The \ct{policy} parameter indicates the kind of operations the unsafe
script can apply on the target node. A \ct{read-only} policy only
grants reading capabilities of the target node to the unsafe script
whereas a \ct{read-write} policy grants full privileges to this node.
Finally, the web developer must include the HTML target node that the
unsafe script will read and update:

\begin{code}
  <div id="unsafeDiv"></div>
\end{code}

\subsubsection*{Step 1. -- Creating the worker}

When \safejs is loaded, it navigates the \dom to find any script
declaration with the \ct{text/safe-javascript} type and creates a
dedicated worker for each. Each worker is initialized with a \safejs{}
\ct{jail.js} script that waits for its configuration by reading its
message queue. No other form of communication is allowed by the worker
draft specification between the main thread and the
worker.

\subsubsection*{Step 2. -- Creating the \vdom}

The message queue only accepts strings. \safejs must then serialize
each message going through the queue: we use
JSON\footnote{\url{http://json.org/}} as the serialization format. The
first message \safejs sends to the worker is a copy of the HTML target
node, \ct{unsafeDiv} in our example. With this message, the worker
creates a \vdom that mimics a real \dom but that can not directly impact
the web document.

\subsubsection*{Step 3. -- Loading the unsafe script}

After the \vdom is created, \safejs sends the URL of the unsafe script
to load (\ct{js/spy.js} in our example) to the worker through the
message queue. When the worker receives this message, it loads the
unsafe script.

\subsubsection*{Step 4. -- Reading from and updating the \vdom}

The unsafe script can use the \vdom as if it was a real \dom. Because
of this, our approach allows any script to be embedded in a web page
with no modification. For the unsafe script, reading and updating the
\vdom is done using the standard \js API. Because the \vdom contains a
(partial) copy of the web page, reading is done without any
communication with the main thread.

\subsubsection*{Step 5. -- Writing to the message queue}

When the unsafe script updates the \vdom, the worker sends a
serialization of the change to the message queue. This change
information contains the state of the \vdom as the unsafe script would
like it to be.

\subsubsection*{Step 6. -- Reading from the message queue}

Then, \safejs deserializes the change and checks the policy to verify
whether the change is allowed. If the change is not allowed, \safejs
can either kill the worker or ignore the update request depending on
its configuration.

\subsubsection*{Step 7. -- Safely updating the \dom}

If the change is allowed, \safejs updates the real \dom. \safejs then
sends a (partial) copy of the \dom to the worker which then updates its
\vdom.

\subsubsection*{Step 8. -- Updating the web document}

When \safejs updates the real \dom, the web browser takes care of
updating the web document.

\section{Architecture of the \vdom}

The \vdom is a full-\js implementation of the \dom\cite{DOM98} level
3. \vdom is actually a fork of the well-known \jsdom \dom
implementation\footnote{\url{https://github.com/tmpvar/jsdom}} that is
used with Node.js by server-side \js
applications.\footnote{\url{http://nodejs.org/}} Our \vdom
implementation is available under a free open-source
license.\footnote{\url{https://gitorious.org/jsdom-client/jsdom-client}}

In this section, we present some of the key differences between \vdom
and \jsdom and explain why they are critical for \safejs.

\subsubsection*{Replace the dependency manager}

The \jsdom library depends on multiple other \js libraries such as
\ct{htmlparser2} and \ct{contextify}. These dependent libraries are
resolved and loaded with a specific instruction that blocks the
execution until the library is ready to be used:

\begin{code}{}
defaultParser = require(htmlparser2)
... code that uses 'defaultParser'...
\end{code}

If this blocking code is perfectly valid for a server-side \js
application, a web browser only allows non-blocking code and there is
no way to force the browser to wait for anything: the web browser
would kill any script with an active waiting loop for example. We thus had to
rewrite all similar code to use an equivalent and non-blocking
function using \requirejs to resolve and load external libraries:

\begin{code}{}
define(["htmlparser2"], function(defaultParser) {
  ... code that uses 'defaultParser'...
})
\end{code}

\paragraph{Remove accesses to the file-system}

\jsdom features the ability to read and write files on the hosting
file system, thanks to a dedicated API provided by \nodejs. Such
unrestricted access to a file system is not possible in a web browser
so we removed this ability in \vdom.

\paragraph{Remove references to the global object}

\jsdom uses \contextify, a C++ library, to bind the pseudo variable
\this to its \window object. In \vdom, we can not depend on a C++
library and we thus had to create a global object that mimics the
behavior of \window. Because \window is, for example, responsible for
providing a logger object (known as the console), the \vdom global
object also provides a virtual logger object.

\paragraph{Recursively apply these changes to \jsdom dependencies}

Finally, we applied the above-mentioned changes to the \jsdom
dependencies we had to integrate in \vdom (\eg cssom, cssstyle and
htmlparser). For this, we leveraged the browserify
tool\footnote{\url{https://github.com/substack/node-browserify}} to

\section{Conclusion}

In this document we reported on the design and implementation of
\safejs, an hermetic sandboxing mechanism for \js. \safejs allows
mashup \js developers to include external (potentially unsafe) scripts
into a web page without risking data leaks from the web page to the
external scripts. Our approach is based on web workers, a draft
specification to run isolated processes in a web page.

The idea behind \safejs is to run each unsafe script inside its own
worker and give this worker a virtual \dom for the script to play
with. This virtual \dom is compatible with the standard \dom but
restricts which changes impact the document.

Our \safejs implementation is currently being used in an industrial
setting in the context of the Resilience FUI 12 project.

\bibliographystyle{alpha}

\begin{thebibliography}{}

\end{thebibliography}


\begin{thebibliography}{ABC{\etalchar{+}}98}

\bibitem[ABC{\etalchar{+}}98]{DOM98}
Vidur Apparao, Steve Byrne, Mike Champion, Scott Isaacs, Ian Jacobs, Arnaud
  Le~Hors, Gavin Nicol, Jonathan Robie, Robert Sutor, Chris Wilson, and Lauren
  Wood.
\newblock {\em Document Object Model Specification {DOM} Level 1 version 1.0}.
\newblock World Wide Web Consortium, 1998.

\bibitem[GSK10]{Guha10a}
Arjun Guha, Claudiu Saftoiu, and Shriram Krishnamurthi.
\newblock The essence of javascript.
\newblock In {\em Proceedings of the 24th European conference on
  Object-oriented programming}, ECOOP'10, pages 126--150, Berlin, Heidelberg,
  2010. Springer-Verlag.

\bibitem[HL07]{Hunt07a}
Galen~C. Hunt and James~R. Larus.
\newblock Singularity: rethinking the software stack.
\newblock {\em SIGOPS Oper. Syst. Rev.}, 41(2):37--49, 2007.

\bibitem[MSL{\etalchar{+}}08]{Mill08a}
Mark~S. Miller, Mike Samuel, Ben Laurie, Ihab Awad, and Mike Stay.
\newblock Caja safe active content in sanitized javascript.
\newblock Technical report, Google Inc., 2008.

\bibitem[MT09]{Maff09a}
Sergio Maffeis and Ankur Taly.
\newblock Language-based isolation of untrusted javascript.
\newblock In {\em In CSF}, 2009.

\bibitem[PEGK11]{Poli11a}
Joe~Gibbs Politz, Spiridon~Aristides Eliopoulos, Arjun Guha, and Shriram
  Krishnamurthi.
\newblock Adsafety: type-based verification of javascript sandboxing.
\newblock In {\em Proceedings of the 20th USENIX conference on Security},
  SEC'11, pages 12--12, Berkeley, CA, USA, 2011. USENIX Association.

\bibitem[RHBV11]{Rich11a}
Gregor Richards, Christian Hammer, Brian Burg, and Jan Vitek.
\newblock The eval that men do: A large-scale study of the use of eval in
  javascript applications.
\newblock In {\em Proceedings of Ecoop 2011}, 2011.

\end{thebibliography}

\newcommand{\etalchar}[1]{$^{#1}$}

\end{document}